\documentclass[twoside,twocolumn,9pt]{article}
\usepackage{extsizes}
\usepackage[super,sort&compress,comma]{natbib} 
\usepackage[version=3]{mhchem}
\usepackage[left=1.5cm, right=1.5cm, top=1.785cm, bottom=2.0cm]{geometry}
\usepackage{balance}
\usepackage{widetext}
\usepackage{times,mathptmx}
\usepackage{sectsty}
\usepackage{graphicx} 
\usepackage{lastpage}
\usepackage[format=plain,justification=raggedright,singlelinecheck=false,font={stretch=1.125,small,sf},labelfont=bf,labelsep=space]{caption}
\usepackage{float}
\usepackage{fancyhdr}
\usepackage{fnpos}
\usepackage[english]{babel}
\usepackage{array}
\usepackage{droidsans}
\usepackage{charter}
\usepackage[T1]{fontenc}
\usepackage[usenames,dvipsnames]{xcolor}
\usepackage{setspace}
\usepackage[compact]{titlesec}
\usepackage{color}
\usepackage{color}

\usepackage{soul} 

\usepackage{epstopdf}

\definecolor{cream}{RGB}{222,217,201}

\begin{document}

\pagestyle{fancy}
\thispagestyle{plain}
\fancypagestyle{plain}{

}




\fancyhead{}
\renewcommand{\headrulewidth}{0pt} 

\makeatletter 
\newlength{\figrulesep} 
\setlength{\figrulesep}{0.5\textfloatsep} 

\newcommand{\topfigrule}{\vspace*{-1pt}%
\noindent{\color{cream}\rule[-\figrulesep]{\columnwidth}{1.5pt}} }

\newcommand{\botfigrule}{\vspace*{-2pt}%
\noindent{\color{cream}\rule[\figrulesep]{\columnwidth}{1.5pt}} }

\newcommand{\dblfigrule}{\vspace*{-1pt}%
\noindent{\color{cream}\rule[-\figrulesep]{\textwidth}{1.5pt}} }

\makeatother

\twocolumn[
  \begin{@twocolumnfalse}

\sffamily
\begin{tabular}{m{2.cm} p{13.5cm} }

 & \noindent\LARGE{\textbf{Dipolar interactions between domains in lipid monolayers at the air-water interface}} \\
%
\vspace{0.3cm} & \vspace{0.3cm} \\

 & \noindent\large{Elena Rufeil-Fiori,$^{\ast}$\textit{$^{a}$} Natalia Wilke,\textit{$^{b}$} and Adolfo J. Banchio\textit{$^{a}$}} \\

 & \noindent\normalsize{

A great variety of biologically relevant monolayers present phase coexistence characterized by domains formed by lipids in an ordered phase state dispersed in a continuous, disordered phase. The difference in surface densities between these phases originates inter--domain dipolar interactions, which are relevant for the determination of the spacial distribution of domains, as well as their dynamics. In this work, we propose a novel manner of estimating the dipolar repulsion using a passive method that involves the analysis of images of the monolayer with phase coexistence. The method is based on the comparison of the pair correlation function obtained from experiments with that obtained from Brownian dynamics simulations of a model system. As an example, we determined the difference in dipolar density of a binary monolayer of DSPC/DMPC at the air--water interface from the analysis of the radial distribution of domains, and the results are compared with those obtained by surface potential determinations. A systematic analysis for experimentally relevant parameter range is given, which may be used as a working curve for obtaining the dipolar repulsion in different systems.


%
%
%

} \\

\end{tabular}

 \end{@twocolumnfalse} \vspace{0.6cm}

  ]

\renewcommand*\rmdefault{bch}\normalfont\upshape
\rmfamily
\section*{}
\vspace{-1cm}

\footnotetext{\textit{$\ast$~E-mail: elena.rufeil@unc.edu.ar}}
\footnotetext{\textit{$^{a}$Instituto de F\'isica Enrique Gaviola, IFEG, CONICET and Facultad de Matem\'atica, Astronom\'ia, F\'isica y Computaci\'on, Universidad Nacional de C\'ordoba, Ciudad Universitaria, X5000HUA, C\'ordoba, Argentina.}}

\footnotetext{\textit{$^{b}$Centro de Investigaciones en Qu\'imica Biol\'ogica de C\'ordoba, CIQUIBIC, CONICET and Departamento de Qu\'imica Biol\'ogica, Facultad de Ciencias Qu\'imicas, Universidad Nacional de C\'ordoba, Ciudad Universitaria, X5000HUA, C\'ordoba, Argentina.}}



\section{Introduction}

In different biomembranes, phase coexistence is frequently observed, depending on the membrane and aqueous composition, temperature and the particular model of membrane (Langmuir monolayers, supported films, free-standing bilayers such as giant unilamellar vesicles or black lipid membranes, among others). When the denser phase is not the continuous phase, domains (named rafts for certain lipid composition) are observed, moving in the more fluid phase. The domains interact with each other~\cite{andelman85, andelman86,ursell09,wilke14}, and these interactions affect their own movement~\cite{wilke14,wilke10} as well as that of other species present in the membrane~\cite{forstner08,ruckerl08}.

Inter-domain interaction may be related with electrostatic forces (dipolar or Coulombic repulsions), those related with the spontaneous curvature of the coexisting phases and hydrodynamic forces that appear when domains are in motion. These forces hinder the coalescence of the domains and modulate the availability of the species in the membrane and their dynamics at long time scales. Dipolar repulsion is always present, since the molecules forming the membrane are ordered and dipolar. Coulombic forces appear for charged domains, while curvature effects are important when the spontaneous curvature of the coexisting phases is markedly different, and for large domains with high line tension~\cite{ursell09}.

The dipolar repulsion may be estimated through the difference in dipole density of the surfactants organized at the interface, including the contribution of the hydration water in the polar head group region~\cite{gawrisch99}. 
A frequently used method for determining the value of the average dipole moment (molecule + hydration water) of a homogeneous film is the determination of the surface potential in Langmuir monolayers at the aqueous/air interface~\cite{gaines, brockman94, taylor}.
Alternatively, probes sensitive to the local potential have been used in bilayers, as well as conductance measurements~\cite{clarke01}. 
Therefore, if the composition of the coexisting phases is known, the dipole potential of monolayers or bilayers with the composition of each phase make it possible to estimate the dipole density of each phase, and thus, the difference between them.

However, the composition of the coexisting phases is not always easily obtained, particularly for systems with more than two components, and the estimation of the difference in dipole density is then not possible from dipole potential measurements. Furthermore, when domains are formed as a consequence of the phase transition of a single--component membrane, the estimation of the dipole potential of each phase at the same temperature and molecular density is not straightforward.

The presence of domains is a common feature in different model membranes and also in plasma membrane of mammals (with nanometer putative sizes)~\cite{pike06}, yeast, fungi and plants (with radii in the micrometer range)~\cite{malinsky13,spira12}. In these natural systems, the composition of the membrane is highly complex and therefore, it is not possible to know the precise composition of the domains, thus preventing also the estimation of difference in dipole density from dipole potential measurements.

Other alternative methods for determining the dipolar repulsion in monolayers with phase coexistence are based
on the analysis of equilibrium size distributions of domains~\cite{mulder03,woog11}. 
The method proposed by Mulder~\cite{mulder03} 
approximates the exact size distribution by a Gaussian and uses a simplified theoretical analysis, where the interdomain interactions are approximately treated.
On the other hand, Lee \textit{et al.}~\cite{woog11} obtain the excess dipolar density by fitting the size distribution with an equilibrium thermodynamic expression. Their scheme assumes no interactions between domains, and hence it is valid for sufficiently diluted domains.

In this work, we propose a novel manner of estimating the dipolar repulsion using a passive method that involves the analysis of images of the monolayer with phase coexistence. 
We make use of the fact that the dipolar repulsion between domains promotes a 2--dimensional spatial arrangement of the domains, in which the average domain--domain distance is maximal. Therefore, the repulsion will induce a domain distribution which leads to the 	radial distribution function characteristic of liquid systems~\cite{hansen}. 
The method is based on the comparison of the pair correlation function obtained from
experiments with Brownian dynamics simulations of a model system.
As an example, we determined the difference in dipolar density of a binary monolayer of 
DSPC/DMPC at the air--water interface from the analysis of the radial distribution of 
domains, and the results are compared with those obtained by surface potential determinations.

\section{Experimental Section}

Mixed distearoylphosphatidylcholine (DSPC) and dimyristoylphosphatidylcholine (DMPC) monolayers present a wide range of compositions and lateral pressures where they exhibit two-phase liquid condensed (LC) and liquid expanded (LE) coexistence region.
At room temperature, the mixed monolayers show LC micrometer sized domains dispersed in a LE continuous phase, at a lateral pressure of 10 mN/m and for composition of DSPC higher than 24 mol$\%$ \cite{wilke10}.

We took micrographs of mixed monolayers at different DSPC concentration using fluorescence microscopy. We perform the experiments with DSPC, DMPC, and the  lipophilic fluorescent probe L-R-phosphatidylethanolamine-N-(lissamine rhodamine B sulfonyl) ammonium salt (chicken egg, trans-phosphatidylated) (Rho-PE) purchased from Avanti Polar Lipids (Alabaster, AL).
The water used for the subphase was from a Milli-Q system (Millipore) with a resistivity of 18 M$\Omega$ cm and a surface tension of 72 mN/m.
 
The fluorescent probe (Rho-PE) was incorporated into the lipid solution before being spread at a concentration of 1 mol$\%$ or less. Monolayers were formed in a Langmuir trough (Microtrough-XS, Kibron Finnland) on subphases of pure water. 
The lipid mixture was dissolved in chloroform:methanol $(2:1)$ to obtain a solution of 1 nmol/$\mu$l, which was spread onto the aqueous surface. 
We performed the experiments at room temperature $T =(20\pm 1 )^{\circ}C$, and films were compressed up to a lateral pressure $\pi=(10\pm 1)$ mN/m determined with a Pt plate using the Wilhelmy method. 

After spreading the lipid layer on an area 1.5 times the lift-off, the subphase level was reduced to a thickness of about 3 mm, in order to minimize convection.
The Langmuir balance was placed on the stage of an inverted microscope (Axiovert 200, Zeiss) equipped with a CCD IxonEM+ model DU-897 (Andor Technology) camera, a 100x objective, a continuous solid state laser (TEM00, 532 nm up to 200 mW, Roithner Lasertech), and rhodamine emission filters. The fluorescent probe partitions preferentially on the LE phase and therefore the domains formed by lipid in a condensed phase appear darker in the images.

The surface potential was determined in Langmuir monolayers using the KSV NIMA Surface Potential Sensor (Helsinki, Finland) with the vibrating plate method. The films were prepared with pure DSPC (composition of the condensed state \cite{wilke10}) or with a mixture of DSPC and DMPC with 24 mole\% of DSPC (composition of the expanded phase \cite{wilke10}).

\subsection{Image analysis and Radial distribution function}

A key quantity to characterize the structure of the monolayer is the
radial distribution function $g(r)$.
Considering an homogeneous distribution of domains in the monolayer plane, 
$g(r)$ represents the probability of finding a domain at the distance $r$ of 
another domain chosen as a reference point:
\begin{equation}
\label{gr}
g(r)=\frac{1}{\rho}  \left\langle \frac{1}{N} \sum_{\substack{i,j=1\\i\neq j}}^N \delta(\vec{r}-\vec{r}_i+\vec{r}_j) \right\rangle \; .
\end{equation}
Here, $\rho = N/A$ is the number density, $N$ the number of domains, $A$ the total monolayer area, $\delta(\vec{r})$ the Dirac delta function and the angular brackets indicate an equilibrium ensemble average.


From the micrographs we calculated the condensed area fraction, $\phi$, 
defined as the ratio of the area occupied by the domains over the monolayer area.
To do this, we determined the amount of each phase, converting the original gray scale images into black/white images using the image processing software {\em ImageJ}~\cite{imageja}. 
Then, the total area occupied by the black regions,
which corresponds to the area occupied by domains, was determined.

To process the images we removed the slightly nonuniform 
illumination in the images 
(due to the intensity distribution across the laser beam profile) using a band-pass filter. Then, we selected a particular gray scale level, and all pixels with intensities above this threshold were converted to ``white'', 
while pixels with intensities below this threshold level were converted to ``black''. 
The value of the threshold level was determined on the basis of an optimal resolution of the structures by performing a constant eye comparison with the original photo. 
The threshold value must be carefully selected, since
it determines the principal source of error for $\phi$~\cite{wilke14}.
Different threshold values change the size of the domains; thus, the determination of $g(r)$ is not significantly affected, since it depends only on the position of the domain centers. However, selection of a low threshold value may lead to an underestimation of the number of domains, since the smallest domains (with a domain area less than 4 pixels)
appear lighter than the larger ones. 
Therefore, the threshold value was selected in order not to modify the total number of domains by more than $10\%$.

We calculate the radial distribution function for each monolayer as a histogram of 
domain center--to--center distance $r$.  
For each condition the order of 1000 micrographs were used of a size of $122 \times 122$ $\mu m^2$, and a binning of 0.5 px (0.12$\mu$m) was selected. This size is larger than the error in $r$, $\Delta r=0.3$ px (0.07$\mu$m), and is small enough to obtain a well--characterized curve.
The error of $g(r)$ is calculated independently for each value of $r$ using
 the standard deviation.

\subsection{Simulations}

We consider a monolayer in its two-phase LC and LE coexistence region, where the LC phase forms domains in the LE phase that occupies the larger area of the monolayer. Because of the difference in surface densities, the LC domains possess an excess dipole density with respect to the surrounding LE phase \cite{mcconnell}. 
This originates dipolar repulsive interactions between the domains.
In general in DSPC--DMPC monolayers, the domains exhibit nearly circular shapes.

\begin{figure}[h]
\centerline{
\includegraphics[width=0.8\columnwidth]{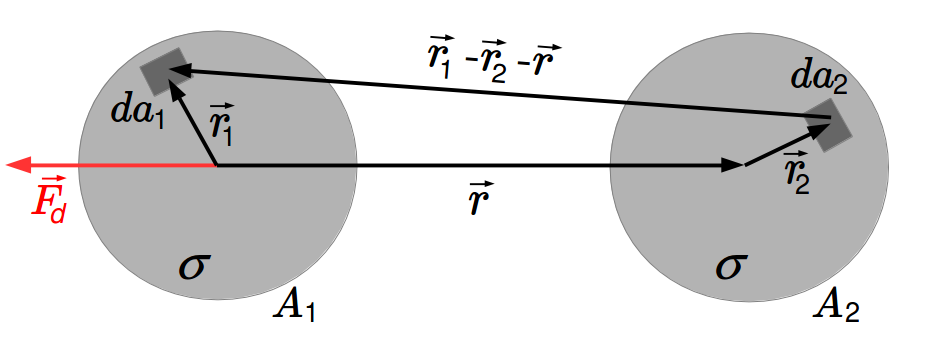} }
\caption{Two domains of equal radii $R$ and excess dipolar density $\sigma$ 
with center--to--center distance $r$.}
\label{esquema}
\end{figure}

We model the mixed monolayer as a uniform layer with permittivity $\epsilon_m$ that lies between two different semi--infinite media (air and water). This layer is composed by monodispersed circular domains with an effective dipole density $\sigma$ perpendicularly oriented to the interface.
Within this model, the resulting dipolar pair potential $U_d$ between two domains can be described by

\begin{equation}
\label{Ud}
 U_d(r)
=\frac{\sigma^2}{4\pi\epsilon_0}\frac{2\epsilon_w\epsilon_a}{\epsilon_m^2(\epsilon_w+\epsilon_a)}\int_{A_2}da_2\int_{A_1}\frac{1}{|\vec{r}_1-\vec{r}_2-\vec{r}|^3}  \; da_1 \;,
\end{equation}
where $A_i$ denotes the area of domain $i$, $da_i$ its area element and $\vec{r}_i$ its position vector respect to the domain center, with $i=1,2$. $\vec{r}$ is the vector from the center of domain 1 to the center of domain 2, as shown in Fig.~\ref{esquema}.
$\epsilon_0$ is the vacuum permittivity, $\epsilon_w$ and $\epsilon_a$ are the relative permittivities of the water and air, respectively. Here, we have used the results from Urbakh \textit{et al.} \cite{urbakh93} that describes the interaction between dipoles in a thin dielectric layer surrounded by two semi--infinite media.

Note that, after defining
\begin{equation}
\label{epss}
 \epsilon^*
=\frac{\epsilon_m^2(\epsilon_w+\epsilon_a)}{2\epsilon_w\epsilon_a},
\end{equation}
the interaction potential $U_d$ results equivalent to that of two domains immersed in an homogeneous medium with an effective permittivity $\varepsilon^*$.

The force on domain 1, derived from this potential is
\begin{equation}
\label{Fd}
 \vec{F}_d(\vec{r}) = 
\frac{3\sigma^2}{4\pi\epsilon_0\epsilon^*}\int_{A_2}da_2\int_{A_1}\frac{\vec{r}_1-\vec{r}_2-\vec{r}}{|\vec{r}_1-\vec{r}_2-\vec{r}|^5}  \; da_1 \;. 
\end{equation}
This force lies along the center--to--center line and its magnitude depends only on the distance $r$ between the domains.
There is no analytic expression for $F_d(r)$, and hence it must be calculated numerically. 
However, for the particular case of monodisperse systems (all domain radii equal to $R$), Wurlitzer \textit{et al.}~\cite{fischer02} found the asymptotic behaviour of this 
force as:
\begin{equation}
\label{Fd_ap}
F_d(r) \approx \left\{ \begin{array}{lcc}
             \frac{\sigma^2}{2\epsilon_0 \epsilon^*}\frac{1}{\sqrt{r/R-2}} &    & 2R < r \ll 3R \\
             \frac{3 \pi \sigma^2}{4\epsilon_0 \epsilon^*} \frac{R^4}{r^4}&   & r \gg 3R \\
             \end{array}
   \right.
\end{equation}
%
As expected, for large distances it reduces to the force of two point dipoles $r^{-4}$.
These expressions do not describe the interaction between domains in the experimentally relevant interval $(2+0.1)R \le r \le 10 R$, and hence a numerical integration of Eq.~(\ref{Fd}) must be performed.
This 4D--integral represent much more computational effort in a simulation than an analytical expression. This is the reason why, it is simpler to approximate the interaction by point dipoles in the center of the domains, with a dipole moment $\mu_i$ representing the dipole density over the area of the domain, $\mu_i=\sigma A_i$.
Then, the pair potential for point dipoles perpendicularly oriented to the interface is
\begin{equation}
\label{Up}
 U_p(r)= \frac{\mu_1 \mu_2}{4\pi\epsilon_0\epsilon^*}\frac{1}{r^3} \;,
\end{equation}
\noindent
and the corresponding force over the domain 1 is
\begin{equation}
\label{Fp}
 \vec{F}_p(\vec{r})= -\frac{3 \mu_1 \mu_2}{4\pi\epsilon_0\epsilon^*}\frac{1}{r^4} \;\hat{r}.
\end{equation}
\noindent
where $\hat{r}$ is ($\hat{r} = \vec{r}/r$) the unitary vector in the direction of $\vec{r}$. 
For convenience, we define
\begin{equation}
\label{f0}
 f_0= \frac{\sigma ^2 }{4\pi\epsilon_0\epsilon^*} \;,
\end{equation}
\noindent
a quantity that characterizes the strength of the forces $F_d$ and $F_p$.

\begin{figure}[h]
\centerline{
\includegraphics[width=0.9\columnwidth]{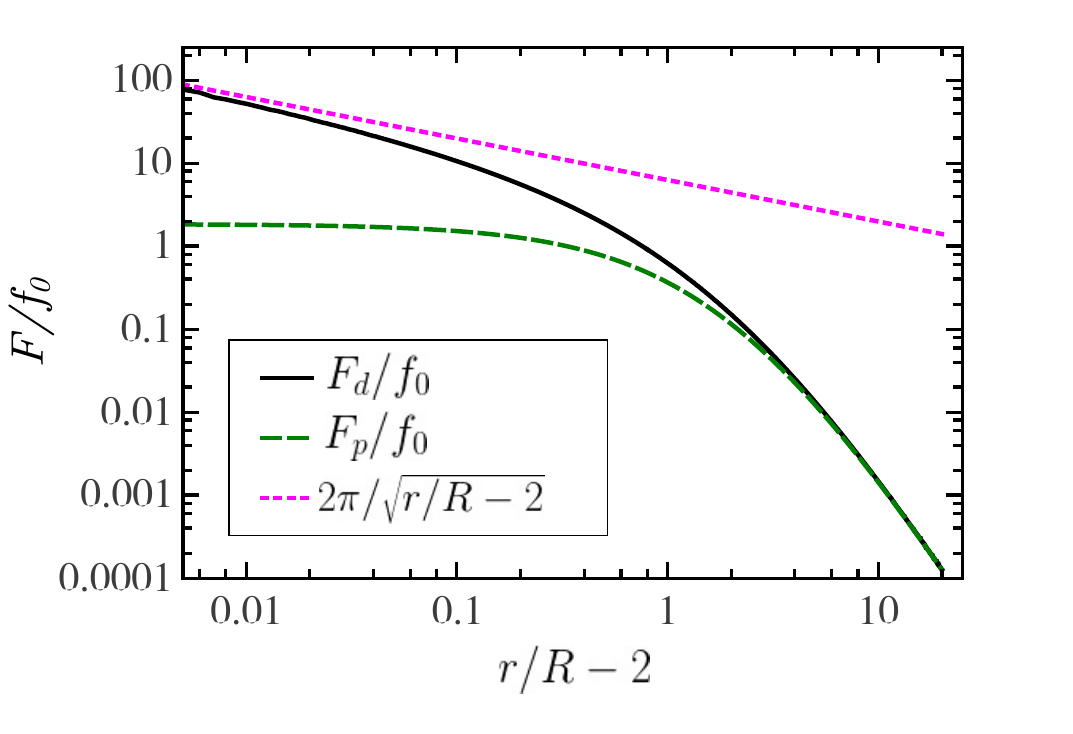}}
\caption{Dipolar interaction force $F_d$ and point dipole approximation force $F_p$ of two domains with equal radii $R$ as a function of the border--to--border separation $r/R-2$, in units of the interaction strength $f_0$. The asymptotic expression of $F_d$ for short distance is also shown.}
\label{fuerzas}
\end{figure}
Figure~\ref{fuerzas} shows $F_d(r)$ (solid line) and $F_p(r)$ (dashed line) for two circular domains of equal radius $R$. Superimposed on the numerical solution of Eq.~(\ref{Fd}), we have added the asymptotic expression for short distance of Eq.~(\ref{Fd_ap}) (dotted line). Note that 
when the two domains approach to contact ($r=2R$) $F_d$ diverges, while $F_p$ 
remains finite.
Furthermore, for the range of experimental interest, $2R < r < 10R$, the difference between the two forces is appreciable. 
In particular, even for $r = 5R$ the difference is of the order of 15\%.

The point dipole approximation deviates from the dipolar density force for short and 
intermediate distances. As a consequence, the structural properties of 
the monolayer obtained with both schemes have remarkable differences.


We model the mixed monolayer as a two dimensional Brownian suspension of interacting 
hard disks of equal radii (monodisperse) immersed in an effective fluid, each disk 
representing an idealized lipid domain.
The inter--domain interactions are described by the full dipolar density pair potential 
given by Eq.~(\ref{Ud}) plus a hard core repulsive part. 
The same system is also studied using the point dipole potential Eq.~(\ref{Up}), in order to analyze the validity of this approximation which is easier to implement.

To study the static properties of the two mixed monolayer models we performed
Brownian dynamics (BD) simulations.
In this scheme, the finite difference equation describing the in-plane displacement of N identical Brownian disks immersed in a fluid during the time step $\Delta t$ is given by \cite{ermak}
\begin{equation}
\label{bd}
 \vec{r}_i(t+\Delta t)- \vec{r}_i(t)=\sum_{j=1}^N \frac{D_0}{k_B T} F_j^P \Delta t + \vec{X}_i  \; ,
\end{equation}
where $F_j^P$ is the direct total force on disk $j$ due to all other N-1 disks, 
$D_0$ the disk diffusion coefficient, $k_B$ the Boltzmann constant, $T$ the temperature,
and $\vec{X}_i$ a random displacement vector of particle $i$ originated from solvent 
particle collisions.
$\vec{X}_i$ is sampled from a Gaussian distribution with zero mean and covariance matrix:
\begin{equation}
\left\langle \vec{X}_i \vec{X}_j \right\rangle=2 D_0  \mathbf{ I}\, \delta_{i,j}\, \Delta t
\end{equation}\noindent
where $\mathbf{I}$ is the identity matrix, and  $\delta_{i,j}$ the Kronecker delta.

The inter--domain dipolar density force, Eq.~(\ref{Fd}),
was calculated in advance, and the values were tabulated for later 
use in the simulations.
The 4d--integral in Eq.~(\ref{Fd}) was calculated using the Monte Carlo algorithm for $2.003 < r/R < 22$, and the analytical expressions for the asymptotes,
Eq.~(\ref{Fd_ap}), were used outside this range.

The simulated systems consisted of $N$ disks of radius $R$ under periodic boundary conditions, using the minimum image convention.
The size of the simulation box, $L$, was determined 
using the expression of the condensed area fraction  $\phi= N \pi R^2/L^2$.

In our simulations, we use $\phi$ determined from the monolayer micrographs, 
$N=144$ disks and $\Delta t=2\times 10^{-4}R^2/D_0$.  
The remaining parameter to determine the system, $f_0$, is systematically varied
in order to find the best agreement with the experimental $g(r)$.

We verified that, for the studied systems, there is no  system size dependency 
in the structural quantities.
The accuracy of our BD simulation method was tested for specific examples by 
comparison with published simulation data on 2D systems \cite{banchio99t}.

\section{Results and Discussions}

\begin{figure}[h!]
\centerline{
\includegraphics[width=0.9\columnwidth]{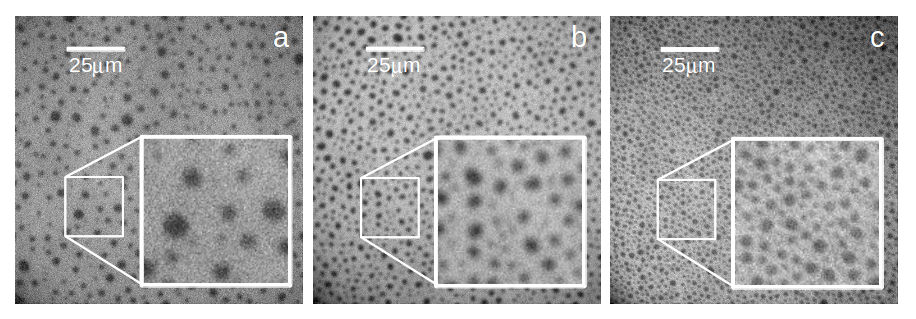}}
\caption{DSPC/DMPC mixed monolayers with (a) 40, (b) 50 and (c) 60 
mol$\%$ DSPC. The amplified zone size is $25 \times 25$ $ \mu m^2$.}
\label{monolayers}
\end{figure}

\begin{figure}[!h]
\centerline{
\includegraphics[width=0.9\columnwidth]{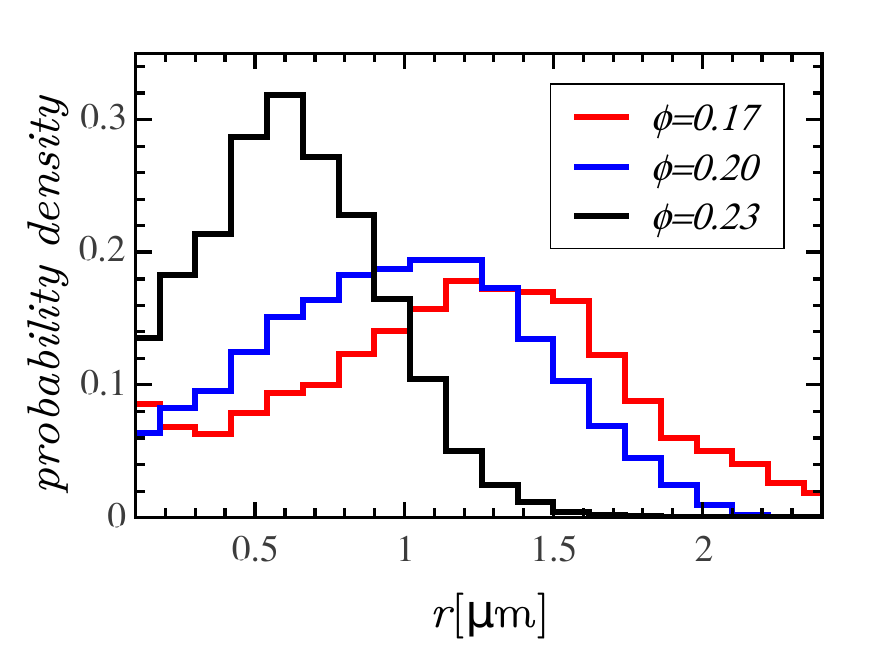}}
\caption{Domain radius distribution of monolayers with different condensed area fraction.}
\label{hist}
\end{figure}

We analyzed monolayers of three DSPC:DMPC compositions with different area fractions.
We used a pure water subphase, 10~mN~m$^{-1}$ lateral pressure, and DSPC mol \%: 40, 50 and 60.
A representative micrograph of each monolayer
is presented in Figure \ref{monolayers}. 
The monolayers were photographed at 12.1~frame/s for 100~s,  while the domains suffered Brownian motion and the monolayer was subjected to drift, enabling different regions of it to be imaged.
The condensed area fraction was calculated for each frame using the process described in the Experimental Section. Then, the value of $\phi$ was obtained from the average.
From the frame area, $a$, and the number of domains in
the micrographs, $n$, the number density  was calculated  as $\rho = n/a$.
The values of $\phi$ and $\rho$ for the systems considered  are shown 
in Table~\ref{tabla1}.
The domain radius distributions were determined from
the domain sizes (areas) assuming circular domains. The corresponding
histograms are shown in Figure~\ref{hist}, and the average radii, $\overline{R}$, 
are presented in Table~\ref{tabla1}.
\begin{table}[!h]
\scriptsize
\centering
\begin{tabular}{|c|c|c|c|c|}
\hline
 DSPC$\pm 2$ & $\phi\pm 0.02$  & $\rho\pm 0.003$  & $\overline{R}$\vphantom{\large X} &   $R_{eff}$  \\
 mol$\%$ &  & $[\mu m^{-2}]$ & $[\mu m]$ & $[\mu m]$   \\ \hline
 40  &  0.17 &  0.028   &   $1.2\pm 0.6$   &   $1.4\pm 0.2$     \\ 
 50  &  0.20 &  0.050   &   $1.0\pm 0.5$   &   $1.1\pm 0.1$     \\ 
 60  &  0.23 &  0.122   &   $0.7\pm 0.3$   &   $0.77\pm 0.07$   \\ 
 \hline
\end{tabular}
\caption{Condensed area fraction, number density, average radius and effective radius for the monolayers considered. Errors are determined by standard deviation, except for $\phi$ which is determined varying the threshold values, as explained in Experimental Section.}
\label{tabla1}
\end{table}

In a monodisperse model the area fraction, the number density and the domain radii are not independent quantities. Since $\phi$ and $\rho$ are more straightforwardly calculated than the average radii, we
determine the effective radius from these quantities as $R_{eff}= ( \phi / (\pi \rho))^{1/2}$.
Note that this radius is slightly different from the average domain radius, due to
the fact that $R_{eff}$ is equivalent to the square root of the second moment of the
radius distribution (see Table~\ref{tabla1}).

\begin{figure}[!h]
\centerline{
\includegraphics[width=0.9\columnwidth]{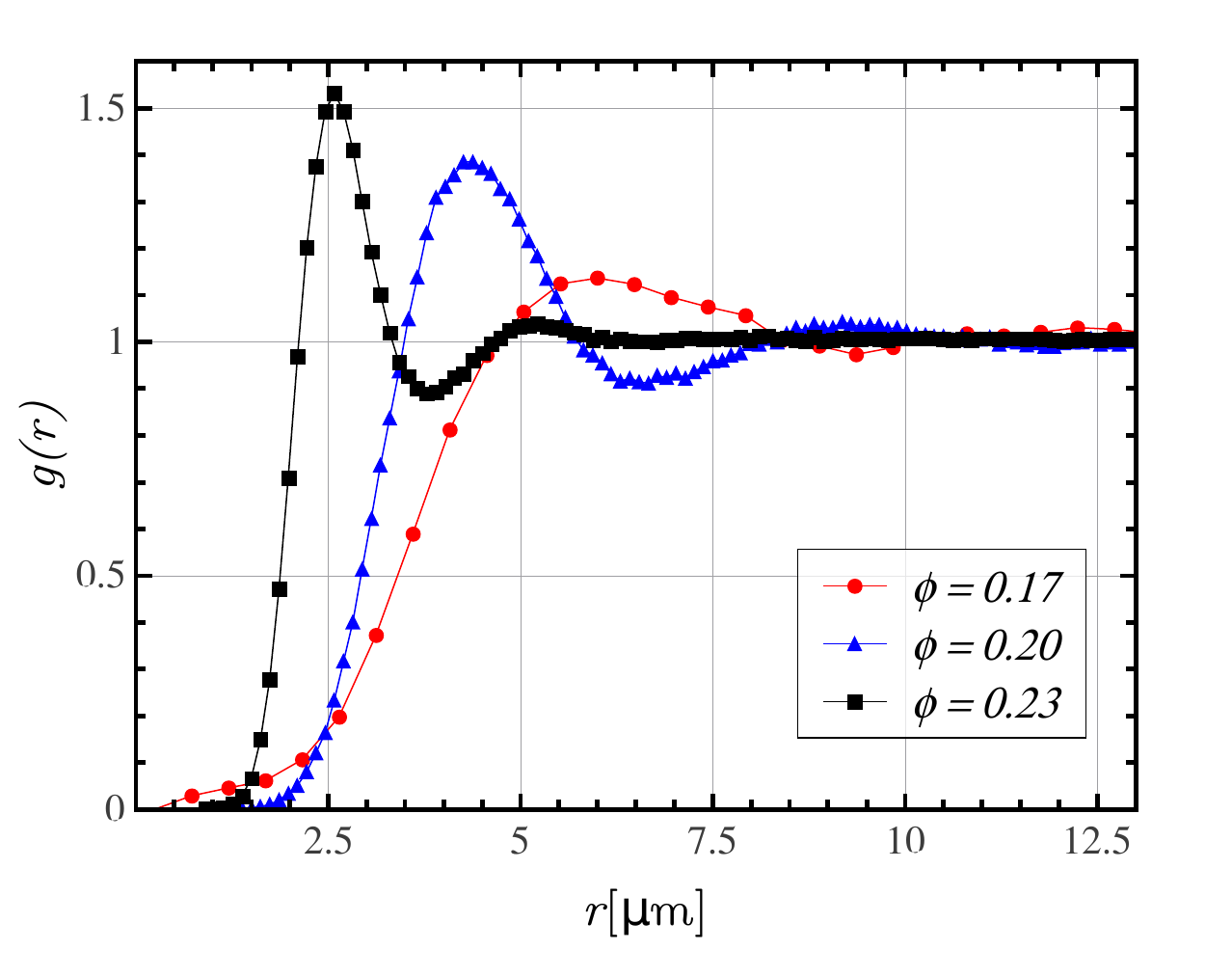}}
\caption{Experimental radial distribution functions, $g(r)$, for the three system analyzed. Error bars are smaller than the symbol size. The lines are guides to the eye.}
\label{gr_exp}
\end{figure}

Figure~\ref{gr_exp} shows the radial distribution functions for the three experimental systems. 
In all cases, the $g(r)$ resembles that of a liquid, showing a well defined first peak
(at $r_{max}$), which characterizes the first coordination shell, i.e., the nearest neighbors.
Besides, a marked first minimum and a second maximum, which corresponds to the second 
coordination shell, is observed for the systems with $\phi=0.20$ and $\phi=0.23$.
Due to the ``strong'' repulsive interaction, the nearest neighbor distance is 
much larger than the average domain diameter.
For this reason the hard core repulsion is not determinant for the structure,
 and the characteristic length of the system is the mean geometric distance 
$r_m = {\rho}^{-1/2}$.

In order to calculate the dipolar repulsion, $\sigma/\sqrt{\epsilon^*}$, 
we fitted the experimental $g(r)$ to the BD one by varying the 
interaction strength $f_0$.
The best fit is selected by matching the height of the first peak. 
Due to the polydispersity in domain sizes, the experimental $g(r)$
has, in general, a broader first peak and a shallower first minimum.
For this reason, we used the first peak height as
fitting criteria instead of fitting the whole curve. 

\begin{figure}[h]
\centerline{
\includegraphics[width=0.9\columnwidth]{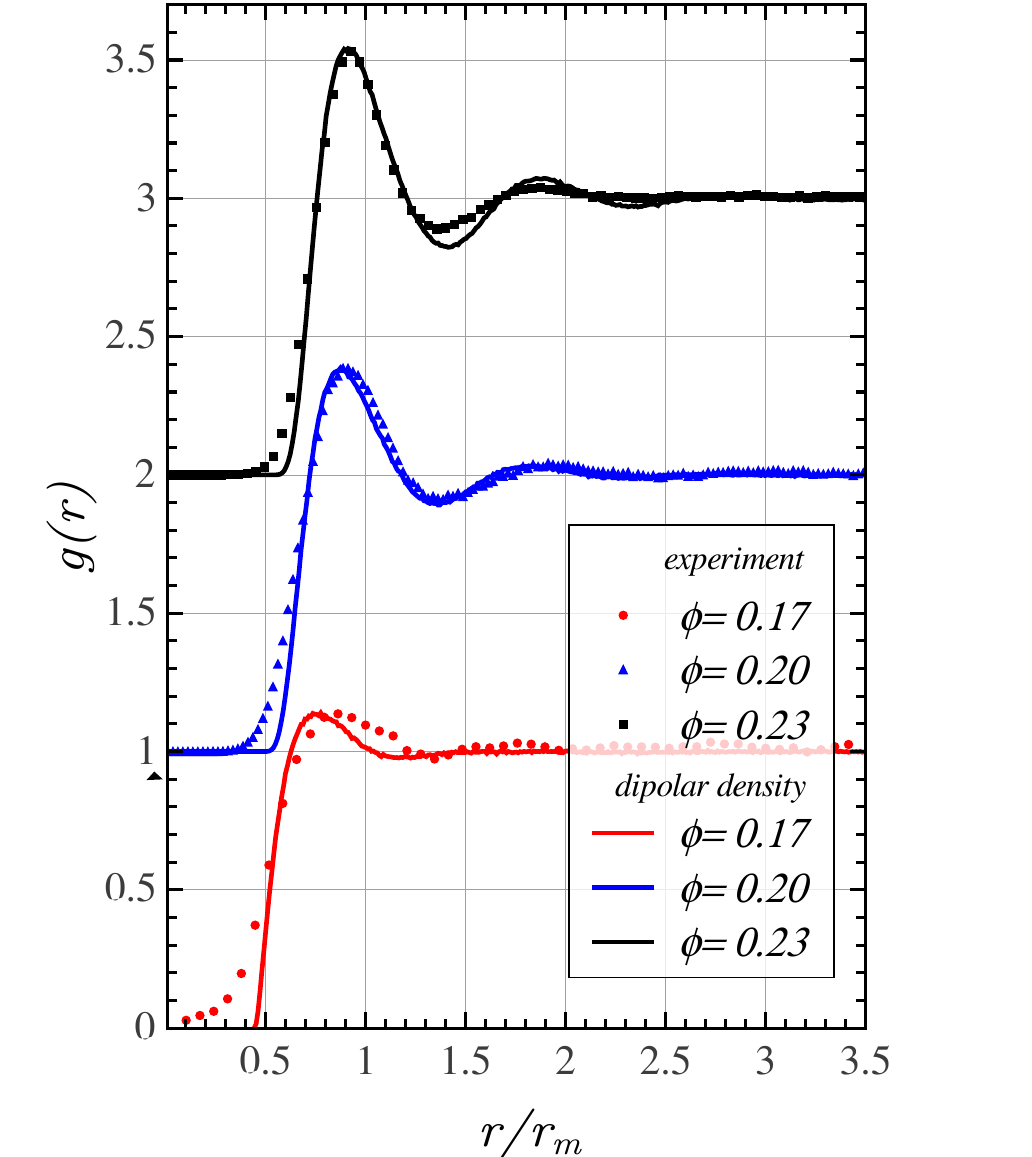}}
\caption{Radial distribution functions obtained experimentally and from BD simulations using dipolar density interactions. For clarity the data have
been vertically shifted.}
\label{gr_exp_sim}
\end{figure}

Figure~\ref{gr_exp_sim} shows the experimental $g(r)$ compared with 
the BD simulations with the full dipolar density interaction, Eq.~\ref{Fd}. 
Here, for convenience,  the distance $r$ is scaled by the mean geometric distance,
$r_m$. 
For the experimental $g(r)$, we selected the value of $r_m$ from the 90\% confidence
 interval that best matches the peak position of the BD $g(r)$. 
In general we observe a remarkable agreement between experiments and simulation.
Simulations results overestimate the first minimum depth and underestimate the
 first peak width.
Both differences might be attributed to polydispersity effects, not taken
into account in our BD simulations.

The fitting procedure, that consists in matching the first peak height at a
fixed $\phi$, has typically a 2\% uncertainty in $f_0$, due to 
the small relative error of the experimental $g(r)$ and the sensitivity of the 
BD $g(r_{max})$ to the interaction strength.
However, the experimental area fraction uncertainty, $\Delta \phi$, needs to be 
considered to determine the error of $f_0$, $\Delta f_0$.
For this reason, we repeated the fitting procedure for two different systems,
with area fractions $\phi \pm \Delta \phi$, respectively. 
From the $f_0$ values obtained, we estimated the contribution to 
$\Delta f_0$ from the uncertainty of $\phi$, $(\Delta f_0)_{\Delta \phi}$, 
which resulted typically of the order of 15\%. 
The BD $g(r)$ obtained as a function of $r/r_m$ are almost indistinguishable from
the one shown in Fig.~\ref{gr_exp_sim}.

In the simulations $f_0$ is used in units of $k_BT/R_{eff}$, 
which result from having $R_{eff}$ as unit of distance and $k_BT$ as unit of energy.
Therefore, the uncertainties of $T$ and $R_{eff}$ must be taken into account
when expressing $f_0$ in physical units. 
However, after error analysis we found that $\Delta f_0$ is dominated only by the
contributions of $(\Delta f_0)_{\Delta \phi}$.

Once we had determined $f_0$ we calculated the dipolar repulsion as:
\begin{equation}
\label{rep}
\frac{\sigma}{\sqrt{\epsilon^*}}=\sqrt{4\pi \epsilon_0 f_0}
\end{equation}
For the systems studied, the values of $\sigma/\sqrt{\epsilon^*}$ are summarized in Table~\ref{tabla2}, where the uncertainties were determined from $\Delta f_0$ discussed above.

In order to compare the dipolar density obtained with other experimental techniques, the value of the relative permittivity of the monolayer $\epsilon_m$ is needed.
In the literature, there are different approaches to describe the permittivity of the monolayer and a wide range of values are found~\cite{andelman86, nassoy, woog11, demchak74, schuhmann89, vogel88, montich85, lelkes80, bockris98, yeh99, bohinc14}.

For this reason, we prefer to use the quantity $\sigma/\epsilon_m$ to compare with other experiments. 
To obtain this quantity within the proposed scheme, we use $\epsilon^*\approx\epsilon_m^2/2$, since $\epsilon_w \gg \epsilon_a$, which leads to $\sigma/\epsilon_m=(0.4\pm 0.1)10^{-12}$C/m, $(0.8\pm 0.1)10^{-12}$C/m and $(1.0\pm 0.1)10^{-12}$C/m for the monolayers with $\phi=0.17$, $0.21$ and $0.23$, respectively.



We performed measurements of the surface potential, $\Delta V$, in two homogeneous monolayers; one in the LE phase and the other in the LC phase, both with the same proportions of DSPC as in the analyzed experiments, as explained in the experimental section.
We obtained $\Delta V_{LC}= ( 555\pm 12 )$ mV and $\Delta V_{LE}= ( 402\pm 8) $ mV, for the pure LC phase and pure LE phase, respectively. 
These values are in agreement with previous results for monolayers with the same~\cite{benvegnu93} and different~\cite{cseh99} subphase, and for bilayers~\cite{starke06}.

Using the parallel plate capacitor model~\cite{gaines}, the excess dipolar density can be obtained from the difference of the surface potential, 
$\Delta(\Delta V)=\Delta V_{LC}-\Delta V_{LE}$, through the equation $\sigma=\Delta(\Delta V)\epsilon_0\epsilon_m$~\cite{schuhmann89}, where the 
same value for the relative permittivity in both phases was assumed. 
The resulting dipolar repulsion obtained from this model is 
$\sigma_c/\epsilon_m=(1.3 \pm 0.2 )10^{-12}$C/m, consistent with the results obtained with the method proposed here.

\begin{table}[!h]
\scriptsize
\centering
\begin{tabular}{|c|c|c|c|}
\hline
 $\phi$ & $f_{0}$ $[k_BT/R_{eff}]$ \vphantom{\large X} & $f_{0}$ $10^{-14}[N]$  & $\sigma/\sqrt{\epsilon^*}$ $10^{-12}[C/m]$ \\ [0.8ex]
 \hline 
0.17\vphantom{\large X} &  $1.2 \pm 0.9$   &  $0.3\pm 0.2$    &  $0.6\pm 0.2$ \\ 
0.20 &  $3 \pm 1$     &  $1.1\pm 0.4$   & $1.1\pm 0.2$  \\ 
0.23 &  $3.7 \pm 0.8$  &  $1.9\pm 0.4$   & $1.5\pm 0.2$ \\ \hline
\end{tabular}
\caption{Interaction strength $f_0$ in simulation units and physical units, and dipolar repulsion $\sigma/\sqrt{\epsilon^*}$ for the analyzed systems.}
\label{tabla2}
\end{table}

We observe that the values obtained for the dipolar repulsion, $\sigma/\sqrt{\epsilon^*}$, 
are indistinguishable, except for the monolayer with $\phi=0.17$.
This could be attributed to the fact that this monolayer has a larger size dispersion and
presents characteristics compatible with a bimodal distribution.
In general, unimodal size polydispersity affects the peak height and width of the 
$g(r)$, while for more complex size distributions these effects are pronounced, and
eventually could also lead to a splitting of the first peak.
In particular, for the monolayer with $\phi=0.17$, these effects are responsible
for the fact that the fitted $g(r)$, obtained with a monodisperse model,
 represents a much less interacting system.
As a consequence, the fitting procedure results in an underestimated value of 
$\sigma/\sqrt{\epsilon^*}$.

\subsection{Radial distribution of domains as a tool for the determination of the inter-domain dipolar repulsion: working curve}

\begin{figure}[h]
\centerline{
\includegraphics[width=0.9\columnwidth]{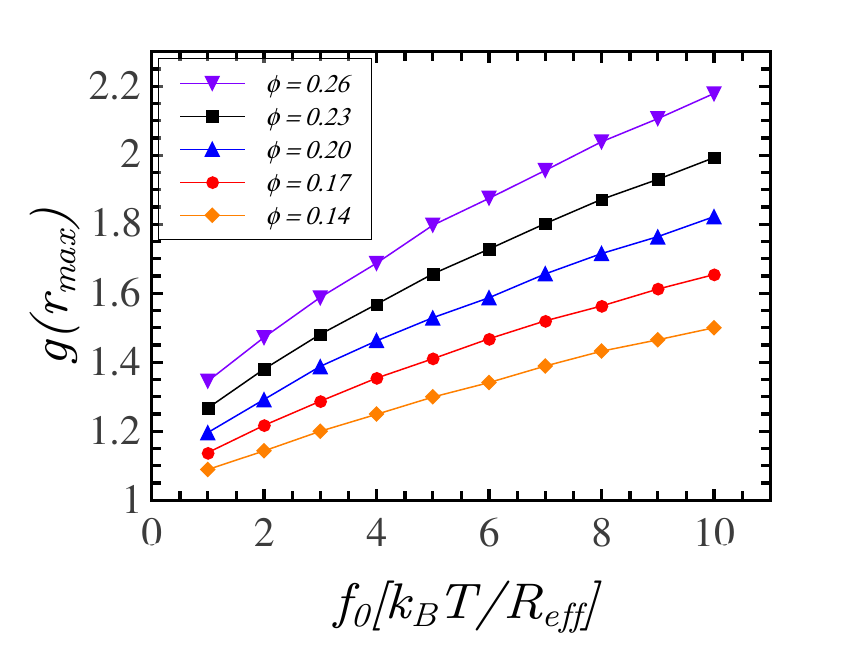}}
\caption{Maximum value of the radial distribution 
function $g(r_{max})$ as function of the interaction strength $f_0$. This relation can be used to obtain the dipolar repulsion $\sigma/\sqrt{\epsilon^*}$ from an experimental value of $g(r_{max})$. Error bars are smaller than the symbol size. The lines are guides to the eye.}
\label{g_r_max}
\end{figure}

With the aim of obtaining the dipolar repulsion for a range of area fraction and interaction strength we performed simulations and analyzed the dependence of $g(r_{max})$ with $\phi$ and $f_0$. Figure~\ref{g_r_max} shows $g(r_{max})$ as a function of $f_0$ in units of $k_BT/R_{eff}$ for five representative values of the area fraction. 

These results may be used to estimate $\sigma/\sqrt{\epsilon^*}$ from the experiments 
without performing simulations. For this purpose, first $\phi$, $\rho$ and $g(r)$ are determined, then $g(r_{max})$ and $R_{eff}$ are calculated, and finally $f_0$ is obtained using Fig.~\ref{g_r_max}.
Note that the value of $\rho$ is used only 
to calculate $R_{eff}$, needed
to express $f_0$ in physical units.
%

\subsection{Point dipole approximation}

\begin{figure}[h]
\centerline{
\includegraphics[width=0.9\columnwidth]{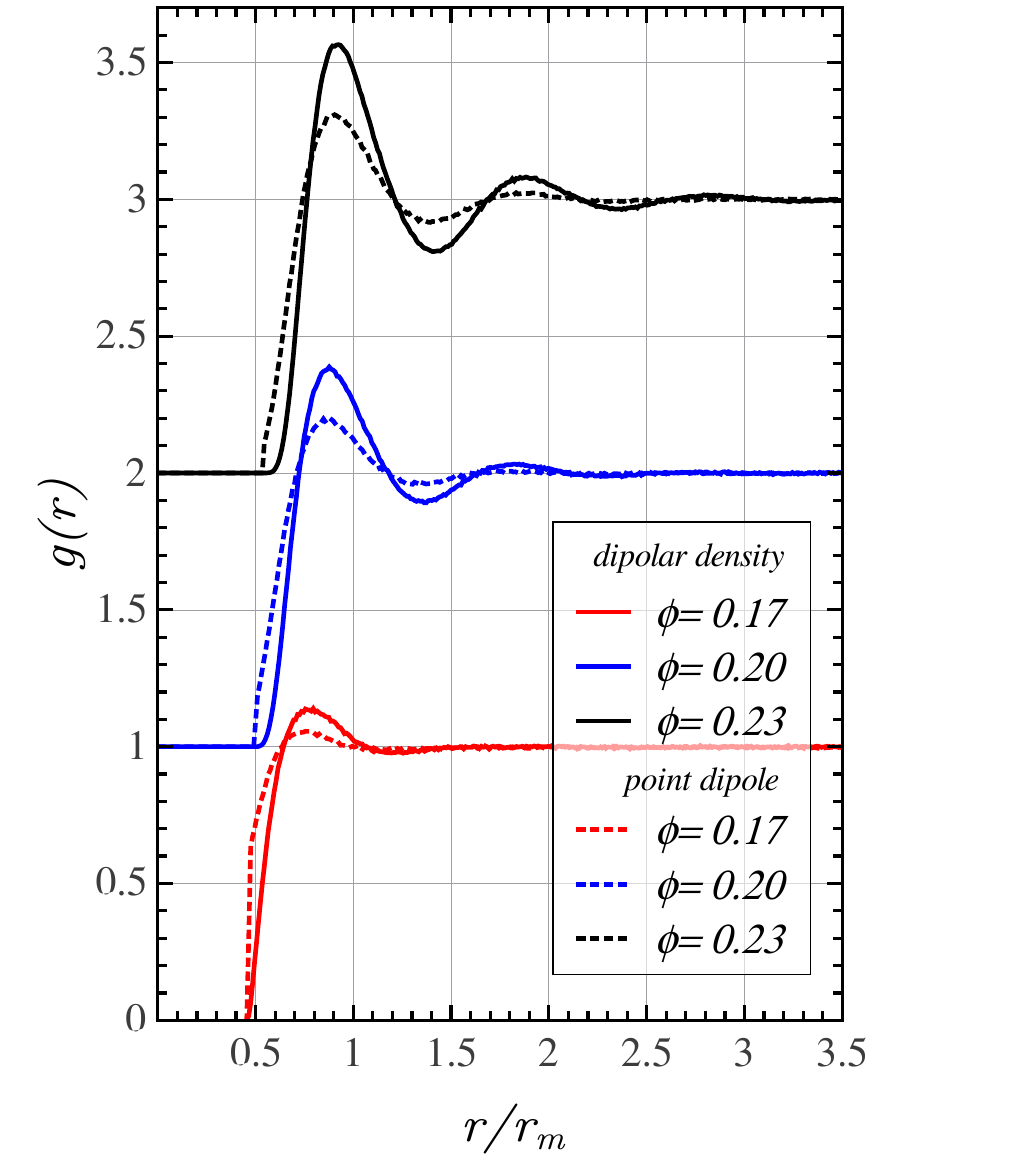}}
\caption{Radial distribution functions obtained from BD simulations 
using dipolar density and point dipole interactions. For each area fraction the 
value of $f_0$ is the same for both models, and is given in Table~\ref{tabla2}.
For clarity the data have been vertically shifted.}
\label{gr_d_p}
\end{figure}

\begin{figure}[h]
\centerline{
\includegraphics[width=0.9\columnwidth]{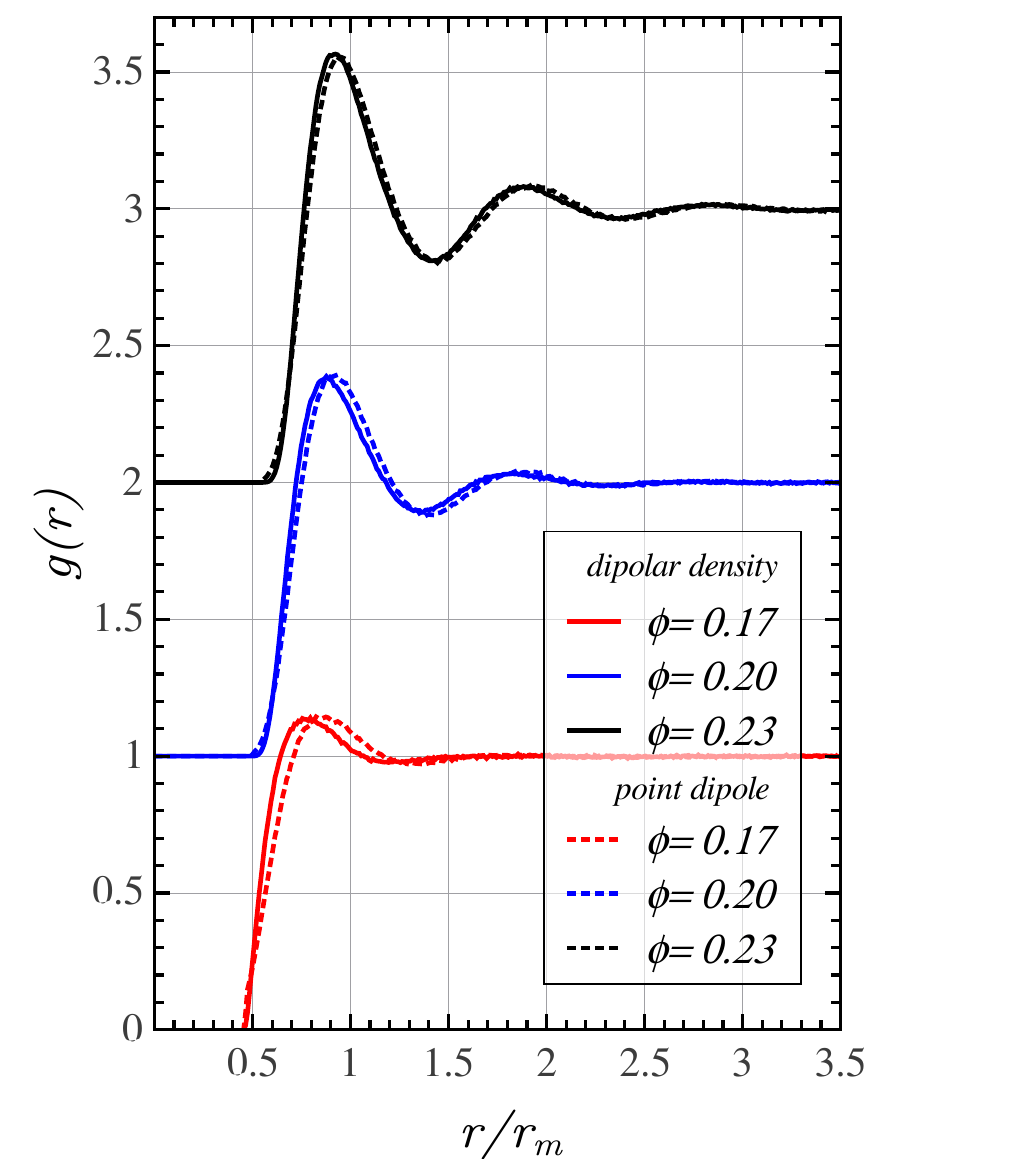}}
\caption{Radial distribution functions obtained  from BD simulations 
using dipolar density and point dipole interactions. 
The point dipole interaction strength $f_0^p$ was selected to match the
respective $g(r_{max})$ of the dipolar density model with parameters 
from Table~\ref{tabla2}. 
The resulting
$f_0^p$ are $3.0$, $6.6$ and $8.2$ $[k_BT/R_{eff}]$ for $\phi=0.17,\, 0.20$ and $0.23$, 
respectively.
%
For clarity the data have been vertically shifted.}
\label{gr_d_p_exp}
\end{figure}

A simple approximation for the domain interactions is
to consider each domain as a point dipole with excluded volume~\cite{forstner08, wilke06}.
This scheme has the advantage of having a simple analytic expression for
the interaction force, Eq.~(\ref{Fp}), hence being easier to implement than
the full dipolar density interaction.
To assess the validity of this approximation in the area fraction range considered,
we systematically studied the structural properties of the point dipole system, and
compared them to the corresponding dipolar density system.
Figure~\ref{gr_d_p} shows $g(r)$ for both schemes using the same set of parameters $[\phi, f_0, \rho]$. It is clearly observed that the point dipole approximation does not  properly represent the structure; in all cases it underestimates the pair correlations.
We found that a good agreement can be achieved if $f_0$ of the point dipole
 model is considered as an effective interaction strength, $f^p_0$, and adjusted to match the first peak height of the dipolar density $g(r)$.
This is shown in Fig.~\ref{gr_d_p_exp} for the three systems analyzed in the 
previous section.
Slight differences are observed in the first peak position and in the small region 
where $g(r)$ starts to deviate from zero.
The agreement improves when the systems are more structured. 
In particular, for weak interacting domains the differences start to be noticeable.
These deviations arise from the fact that the point dipole force tends to a finite
value at contact, while in the dipolar density case it diverges.
For this reason, the hard disk interaction becomes relevant for the point dipole 
model.

As expected $f^p_0$ is always larger than $f_0$. 
Therefore, even if a good agreement is achieved, fitting the experimental data with 
the point dipole model leads to an overestimated value of the dipolar repulsion.
For the monolayer with $\phi=0.20$, fitted with $f^p_0 = 6.6 R_{eff}/(k_BT)$, a
dipolar repulsion $\sigma/\sqrt{\epsilon^*} = 1.6\times 10^{-12}C/m$ would result, 
which is $45\%$ larger than the one calculated with the full dipolar density model 
(see Table~\ref{tabla2}). 

\begin{figure}[h]
\centerline{
\includegraphics[width=0.90\columnwidth]{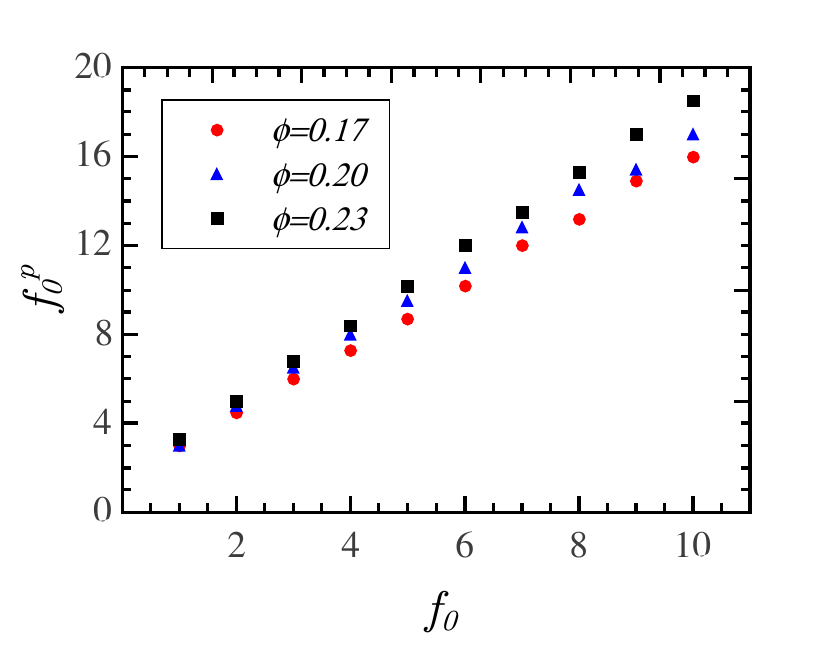}}
\caption{Effective interaction strength $f_0^p$ for the point dipole approximation versus the interaction strength $f_0$. Error bars are smaller than the symbol size.}
\label{f0p_f0d}
\end{figure}

Due to the simplicity of the point dipole model, it is worth looking for a correspondence between both models. 
With this purpose, we determined the effective interaction strength, $f^p_0$,
for a set of $f_0$ at three different area fractions. 
For the range of parameters of interest the
relation between $f^p_0$ and $f_0$ can be seen in Fig.~\ref{f0p_f0d}.
This figure can be helpful for selecting the appropriate $f^p_0$ in order to compute the
structural properties of a monolayer by performing simulations based only on
the simpler point dipole model.
Alternatively, if the experimental pair correlation function is modeled by the
point dipole approximation, the value of $\sigma/\sqrt{\epsilon^*}$ can be calculated using the
corresponding $f_0$ estimated from this figure.

\section{Conclusions}

In the present work, the inter--domain dipolar repulsion in a mixed monolayer was
calculated from the experimental radial distribution functions using Brownian 
dynamics simulations.
The domains were modeled as monodisperse disks with constant dipolar density, 
and their pair interaction obtained from Monte Carlo integration.
We found good  agreement between experiments and simulations for the 
structure of the monolayers, where $\sigma/\sqrt{\epsilon^*}$ was used as a fitting 
parameter.
Mixed monolayers with three different condensed area fraction were analyzed.

With the exception of the experiment with $\phi=0.17$, the  values obtained of $\sigma/\sqrt{\epsilon^*}$ were indistinguishable.  This finding is in agreement with the fact that the studied monolayers are located in the regime of the condensed area fraction where the lever rule is valid~\cite{wilke10}, and hence no $\phi$--dependence of $\sigma$ is expected. 
Furthermore, the dipolar repulsion obtained is of the same order of magnitude as that obtained with the parallel plate capacitor model. 

The dipolar repulsion obtained from the experiment with $\phi=0.17$ has a lower value.
This can be attributed to the fact that this monolayer has a broad and non--unimodal 
size distribution, whose effects are not taken into account in the simulations.
Here, for simplicity, we have modeled the domains as monodisperse disks, 
while experimental monolayers show a distribution of sizes.
It is known that this assumption induces a ``softening'' of the pair correlation 
function due to averaging over different particle sizes\cite{pond11}.
These effects are expected to be relevant specially when the size distribution presents 
a complex structure or has large variance.
The inclusion of polydispersity is feasible in our scheme, and is left for future work.

We have systematically studied the dependence of $g(r_{max})$ with the interaction
strength and area fraction. 
The results, summarized in Figure~\ref{g_r_max}, can be used as a working curve to
estimate $\sigma/\sqrt{\epsilon^*}$ directly from the experimental data without implementing 
any simulation, for the range of parameters studied.

We have also studied the point dipole model to probe its validity as an approximation
to the dipolar density model.
Analyzing the radial distribution function, we found that the structure 
is strongly underestimated by the point dipole model.
However, if the point dipole strength is considered as a fitting parameter, i.e.
as an effective interaction strength, a good agreement can be achieved.
For different area fractions we determined $f_0^p$ for a wide range of $f_0$. 
This relation between the models 
allows the use of the simpler point dipole model both, to calculate the
structure for a monolayer with known dipolar repulsion and to obtain the
dipolar repulsion from an experimental $g(r)$.

We conclude that the proposed procedure to estimate $\sigma/\sqrt{\epsilon^*}$ from the 
experimental pair correlation function constitutes an alternative to conventional 
methods, with the advantage that it requires only monolayer micrographs.

The proposed method may be used not only for binary lipid monolayers, but also for single--component monolayers and more complex monolayers (such as those prepared from isolated membranes, i.e. with all the original components of the cell membrane), and also for bilayers, when domains are present only in one hemi--layer and the hemi--layers are uncoupled or loosely coupled,
provided that a time--averaged radial distribution can be obtained, and that the predominating interactions corresponds to those derived from dipolar repulsion.
Other functional forms may also be explored in a similar manner to that shown here.

\subsection*{Acknowledgement}
The authors acknowledge financial support from CONICET and SeCyT-UNC. ERF thanks S. Ceppi for fruitful discussions.

%


\bibliography{mr_bib_new} 

\providecommand*{\mcitethebibliography}{\thebibliography}
\csname @ifundefined\endcsname{endmcitethebibliography}
{\let\endmcitethebibliography\endthebibliography}{}
\begin{mcitethebibliography}{38}
\providecommand*{\natexlab}[1]{#1}
\providecommand*{\mciteSetBstSublistMode}[1]{}
\providecommand*{\mciteSetBstMaxWidthForm}[2]{}
\providecommand*{\mciteBstWouldAddEndPuncttrue}
  {\def\EndOfBibitem{\unskip.}}
\providecommand*{\mciteBstWouldAddEndPunctfalse}
  {\let\EndOfBibitem\relax}
\providecommand*{\mciteSetBstMidEndSepPunct}[3]{}
\providecommand*{\mciteSetBstSublistLabelBeginEnd}[3]{}
\providecommand*{\EndOfBibitem}{}
\mciteSetBstSublistMode{f}
\mciteSetBstMaxWidthForm{subitem}
{(\emph{\alph{mcitesubitemcount}})}
\mciteSetBstSublistLabelBeginEnd{\mcitemaxwidthsubitemform\space}
{\relax}{\relax}

\bibitem[Andelman \emph{et~al.}(1985)Andelman, Brochard, de~Gennes, and
  Joanny]{andelman85}
D.~Andelman, F.~Brochard, P.~G. de~Gennes and J.~F. Joanny, \emph{C. R. Acad.
  Sc. Paris}, 1985, \textbf{301}, 675--679\relax
\mciteBstWouldAddEndPuncttrue
\mciteSetBstMidEndSepPunct{\mcitedefaultmidpunct}
{\mcitedefaultendpunct}{\mcitedefaultseppunct}\relax
\EndOfBibitem
\bibitem[Andelman \emph{et~al.}(1986)Andelman, Brochard, and
  Joanny]{andelman86}
D.~Andelman, F.~Brochard and J.~F. Joanny, \emph{J. Chem. Phys.}, 1986,
  \textbf{86}, 3673--3681\relax
\mciteBstWouldAddEndPuncttrue
\mciteSetBstMidEndSepPunct{\mcitedefaultmidpunct}
{\mcitedefaultendpunct}{\mcitedefaultseppunct}\relax
\EndOfBibitem
\bibitem[Ursell and W.~S.~Klug(2009)]{ursell09}
T.~S. Ursell and R.~P. W.~S.~Klug, \emph{Proc. Natl. Acad. Sci.}, 2009,
  \textbf{106}, 13301--13306\relax
\mciteBstWouldAddEndPuncttrue
\mciteSetBstMidEndSepPunct{\mcitedefaultmidpunct}
{\mcitedefaultendpunct}{\mcitedefaultseppunct}\relax
\EndOfBibitem
\bibitem[Caruso \emph{et~al.}(2014)Caruso, Villareal, Reinaudi, and
  Wilke]{wilke14}
B.~Caruso, M.~Villareal, L.~Reinaudi and N.~Wilke, \emph{J. Phys. Chem. B},
  2014, \textbf{118}, 519--529\relax
\mciteBstWouldAddEndPuncttrue
\mciteSetBstMidEndSepPunct{\mcitedefaultmidpunct}
{\mcitedefaultendpunct}{\mcitedefaultseppunct}\relax
\EndOfBibitem
\bibitem[Wilke \emph{et~al.}(2010)Wilke, Vega~Mercado, and Maggio]{wilke10}
N.~Wilke, F.~Vega~Mercado and B.~Maggio, \emph{Langmuir}, 2010, \textbf{26},
  11050--11059\relax
\mciteBstWouldAddEndPuncttrue
\mciteSetBstMidEndSepPunct{\mcitedefaultmidpunct}
{\mcitedefaultendpunct}{\mcitedefaultseppunct}\relax
\EndOfBibitem
\bibitem[Forstner \emph{et~al.}(2008)Forstner, Martin, R{\"u}ckerl, K{\"a}s,
  and Selle]{forstner08}
M.~Forstner, D.~Martin, F.~R{\"u}ckerl, J.~A. K{\"a}s and C.~Selle, \emph{Phys.
  Rev. E}, 2008, \textbf{77}, 051906--1--7\relax
\mciteBstWouldAddEndPuncttrue
\mciteSetBstMidEndSepPunct{\mcitedefaultmidpunct}
{\mcitedefaultendpunct}{\mcitedefaultseppunct}\relax
\EndOfBibitem
\bibitem[R{\"u}kerl \emph{et~al.}(2008)R{\"u}kerl, K{\"a}s, and
  Selle]{ruckerl08}
F.~R{\"u}kerl, J.~A. K{\"a}s and C.~Selle, \emph{Langmuir}, 2008, \textbf{24},
  3365--3369\relax
\mciteBstWouldAddEndPuncttrue
\mciteSetBstMidEndSepPunct{\mcitedefaultmidpunct}
{\mcitedefaultendpunct}{\mcitedefaultseppunct}\relax
\EndOfBibitem
\bibitem[Gawrisch \emph{et~al.}(1992)Gawrisch, Ruston, Zimmerberg, Parsegian,
  Rand, and Fuller]{gawrisch99}
K.~Gawrisch, D.~Ruston, J.~Zimmerberg, V.~Parsegian, R.~Rand and N.~Fuller,
  \emph{Biophys. J.}, 1992, \textbf{61}, 1213--1223\relax
\mciteBstWouldAddEndPuncttrue
\mciteSetBstMidEndSepPunct{\mcitedefaultmidpunct}
{\mcitedefaultendpunct}{\mcitedefaultseppunct}\relax
\EndOfBibitem
\bibitem[Gaines(1966)]{gaines}
G.~L. Gaines, \emph{Insoluble Monolayers at Liquid-Gas Interfaces},
  Interscience Publishers, New York, 1966\relax
\mciteBstWouldAddEndPuncttrue
\mciteSetBstMidEndSepPunct{\mcitedefaultmidpunct}
{\mcitedefaultendpunct}{\mcitedefaultseppunct}\relax
\EndOfBibitem
\bibitem[Brockman(1994)]{brockman94}
H.~L. Brockman, \emph{Chem. Phys. Lipids}, 1994, \textbf{73}, 57--79\relax
\mciteBstWouldAddEndPuncttrue
\mciteSetBstMidEndSepPunct{\mcitedefaultmidpunct}
{\mcitedefaultendpunct}{\mcitedefaultseppunct}\relax
\EndOfBibitem
\bibitem[Taylor(2000)]{taylor}
D.~M. Taylor, \emph{Adv. Colloid Interface Sci.}, 2000, \textbf{87},
  183--203\relax
\mciteBstWouldAddEndPuncttrue
\mciteSetBstMidEndSepPunct{\mcitedefaultmidpunct}
{\mcitedefaultendpunct}{\mcitedefaultseppunct}\relax
\EndOfBibitem
\bibitem[Clarke(2001)]{clarke01}
R.~J. Clarke, \emph{Adv. Colloid Interface Sci.}, 2001, \textbf{89},
  263--281\relax
\mciteBstWouldAddEndPuncttrue
\mciteSetBstMidEndSepPunct{\mcitedefaultmidpunct}
{\mcitedefaultendpunct}{\mcitedefaultseppunct}\relax
\EndOfBibitem
\bibitem[Pike(2006)]{pike06}
L.~Pike, \emph{J. Lipid. Res.}, 2006, \textbf{47}, 1597--1598\relax
\mciteBstWouldAddEndPuncttrue
\mciteSetBstMidEndSepPunct{\mcitedefaultmidpunct}
{\mcitedefaultendpunct}{\mcitedefaultseppunct}\relax
\EndOfBibitem
\bibitem[Malinsky \emph{et~al.}(2013)Malinsky, Opekarova, Grossmann, and
  Tanner]{malinsky13}
J.~Malinsky, M.~Opekarova, G.~Grossmann and W.~Tanner, \emph{Annu. Rev. Plant.
  Biol.}, 2013, \textbf{64}, 501--529\relax
\mciteBstWouldAddEndPuncttrue
\mciteSetBstMidEndSepPunct{\mcitedefaultmidpunct}
{\mcitedefaultendpunct}{\mcitedefaultseppunct}\relax
\EndOfBibitem
\bibitem[Spira \emph{et~al.}(2012)Spira, Mueller, Beck, von Olshausen, Beig,
  and Wedlich-S{\"o}ldner]{spira12}
F.~Spira, N.~S. Mueller, G.~Beck, P.~von Olshausen, J.~Beig and
  R.~Wedlich-S{\"o}ldner, \emph{Nat. Cell Biol.}, 2012, \textbf{14},
  640--648\relax
\mciteBstWouldAddEndPuncttrue
\mciteSetBstMidEndSepPunct{\mcitedefaultmidpunct}
{\mcitedefaultendpunct}{\mcitedefaultseppunct}\relax
\EndOfBibitem
\bibitem[Mulder(2003)]{mulder03}
W.~H. Mulder, \emph{J. Colloid Interface Sci.}, 2003, \textbf{264},
  558–560\relax
\mciteBstWouldAddEndPuncttrue
\mciteSetBstMidEndSepPunct{\mcitedefaultmidpunct}
{\mcitedefaultendpunct}{\mcitedefaultseppunct}\relax
\EndOfBibitem
\bibitem[Lee \emph{et~al.}(2011)Lee, Y.~Min, Ramachandran, Iseaelachvili, and
  Zasadzinski]{woog11}
D.~W. Lee, P.~D. Y.~Min, A.~Ramachandran, J.~N. Iseaelachvili and J.~A.
  Zasadzinski, \emph{Proceedings of the National Academy of Sciences}, 2011,
  \textbf{108}, 9425--9430\relax
\mciteBstWouldAddEndPuncttrue
\mciteSetBstMidEndSepPunct{\mcitedefaultmidpunct}
{\mcitedefaultendpunct}{\mcitedefaultseppunct}\relax
\EndOfBibitem
\bibitem[Hansen and McDonald(2013)]{hansen}
J.~P. Hansen and I.~R. McDonald, \emph{Theory of Simple Liquids}, Academic
  Press, Oxford, 4th edn, 2013\relax
\mciteBstWouldAddEndPuncttrue
\mciteSetBstMidEndSepPunct{\mcitedefaultmidpunct}
{\mcitedefaultendpunct}{\mcitedefaultseppunct}\relax
\EndOfBibitem
\bibitem[Schneider \emph{et~al.}(2012)Schneider, Rasband, and
  Eliceiri]{imageja}
C.~A. Schneider, W.~S. Rasband and K.~W. Eliceiri, \emph{Nature Methods}, 2012,
  \textbf{9}, 671--675\relax
\mciteBstWouldAddEndPuncttrue
\mciteSetBstMidEndSepPunct{\mcitedefaultmidpunct}
{\mcitedefaultendpunct}{\mcitedefaultseppunct}\relax
\EndOfBibitem
\bibitem[McConnell(1991)]{mcconnell}
H.~M. McConnell, \emph{Annu. Rev. Phys. Chem.}, 1991, \textbf{42},
  171--195\relax
\mciteBstWouldAddEndPuncttrue
\mciteSetBstMidEndSepPunct{\mcitedefaultmidpunct}
{\mcitedefaultendpunct}{\mcitedefaultseppunct}\relax
\EndOfBibitem
\bibitem[Urbakh and Klafter(1993)]{urbakh93}
M.~Urbakh and J.~Klafter, \emph{J. Phys. Chem.}, 1993, \textbf{97},
  3344--3349\relax
\mciteBstWouldAddEndPuncttrue
\mciteSetBstMidEndSepPunct{\mcitedefaultmidpunct}
{\mcitedefaultendpunct}{\mcitedefaultseppunct}\relax
\EndOfBibitem
\bibitem[Wurlitzer \emph{et~al.}(2002)Wurlitzer, Schmiedel, and
  Fischer]{fischer02}
S.~Wurlitzer, H.~Schmiedel and T.~M. Fischer, \emph{Langmuir}, 2002,
  \textbf{18}, 4393--4400\relax
\mciteBstWouldAddEndPuncttrue
\mciteSetBstMidEndSepPunct{\mcitedefaultmidpunct}
{\mcitedefaultendpunct}{\mcitedefaultseppunct}\relax
\EndOfBibitem
\bibitem[Ermak and McCammon(1978)]{ermak}
D.~Ermak and J.~A. McCammon, \emph{J. Chem. Phys.}, 1978, \textbf{69},
  1352--1360\relax
\mciteBstWouldAddEndPuncttrue
\mciteSetBstMidEndSepPunct{\mcitedefaultmidpunct}
{\mcitedefaultendpunct}{\mcitedefaultseppunct}\relax
\EndOfBibitem
\bibitem[Banchio(1999)]{banchio99t}
A.~J. Banchio, \emph{PhD thesis}, Universit{\"a}t Konstanz, Konstanz, Germany,
  1999\relax
\mciteBstWouldAddEndPuncttrue
\mciteSetBstMidEndSepPunct{\mcitedefaultmidpunct}
{\mcitedefaultendpunct}{\mcitedefaultseppunct}\relax
\EndOfBibitem
\bibitem[Nassoy \emph{et~al.}(1996)Nassoy, Birch, Andelman, and
  Rondelez]{nassoy}
P.~Nassoy, W.~R. Birch, D.~Andelman and F.~Rondelez, \emph{Phys. Rev. Lett.},
  1996, \textbf{76}, 455--458\relax
\mciteBstWouldAddEndPuncttrue
\mciteSetBstMidEndSepPunct{\mcitedefaultmidpunct}
{\mcitedefaultendpunct}{\mcitedefaultseppunct}\relax
\EndOfBibitem
\bibitem[Demchak and Jr.(1974)]{demchak74}
R.~J. Demchak and T.~F. Jr., \emph{J. Colloid and Interface Science}, 1974,
  \textbf{46}, 191--202\relax
\mciteBstWouldAddEndPuncttrue
\mciteSetBstMidEndSepPunct{\mcitedefaultmidpunct}
{\mcitedefaultendpunct}{\mcitedefaultseppunct}\relax
\EndOfBibitem
\bibitem[Schuhmann(1989)]{schuhmann89}
D.~Schuhmann, \emph{J. Colloid Interface Sci.}, 1989, \textbf{134},
  152--160\relax
\mciteBstWouldAddEndPuncttrue
\mciteSetBstMidEndSepPunct{\mcitedefaultmidpunct}
{\mcitedefaultendpunct}{\mcitedefaultseppunct}\relax
\EndOfBibitem
\bibitem[Vogel and M{\"o}bius(1988)]{vogel88}
V.~Vogel and D.~M{\"o}bius, \emph{J. Colloid and Interface Science}, 1988,
  \textbf{126}, 408--420\relax
\mciteBstWouldAddEndPuncttrue
\mciteSetBstMidEndSepPunct{\mcitedefaultmidpunct}
{\mcitedefaultendpunct}{\mcitedefaultseppunct}\relax
\EndOfBibitem
\bibitem[Montich \emph{et~al.}(1985)Montich, Bustos, Maggio, and
  Cumar]{montich85}
G.~G. Montich, M.~M. Bustos, B.~Maggio and F.~A. Cumar, \emph{Chemistry and
  Physics of Lipids}, 1985, \textbf{38}, 319--326\relax
\mciteBstWouldAddEndPuncttrue
\mciteSetBstMidEndSepPunct{\mcitedefaultmidpunct}
{\mcitedefaultendpunct}{\mcitedefaultseppunct}\relax
\EndOfBibitem
\bibitem[Lelkes and Miller(1980)]{lelkes80}
P.~I. Lelkes and I.~R. Miller, \emph{J. Membrane Biol.}, 1980, \textbf{52},
  1--15\relax
\mciteBstWouldAddEndPuncttrue
\mciteSetBstMidEndSepPunct{\mcitedefaultmidpunct}
{\mcitedefaultendpunct}{\mcitedefaultseppunct}\relax
\EndOfBibitem
\bibitem[Bockris and Reddy(1998)]{bockris98}
J.~O. Bockris and A.~K.~N. Reddy, \emph{Modern Electrochemistry}, Plenum Press,
  2nd edn, 1998\relax
\mciteBstWouldAddEndPuncttrue
\mciteSetBstMidEndSepPunct{\mcitedefaultmidpunct}
{\mcitedefaultendpunct}{\mcitedefaultseppunct}\relax
\EndOfBibitem
\bibitem[Yeh and Berkowitza(1999)]{yeh99}
I.~C. Yeh and M.~L. Berkowitza, \emph{J. Chem. Phys.}, 1999, \textbf{110},
  7935--7942\relax
\mciteBstWouldAddEndPuncttrue
\mciteSetBstMidEndSepPunct{\mcitedefaultmidpunct}
{\mcitedefaultendpunct}{\mcitedefaultseppunct}\relax
\EndOfBibitem
\bibitem[Bohinc \emph{et~al.}(2014)Bohinc, Giner-Casares, and May]{bohinc14}
K.~Bohinc, J.~J. Giner-Casares and S.~May, \emph{J. Phys. Chem. B}, 2014,
  \textbf{118}, 7568--7576\relax
\mciteBstWouldAddEndPuncttrue
\mciteSetBstMidEndSepPunct{\mcitedefaultmidpunct}
{\mcitedefaultendpunct}{\mcitedefaultseppunct}\relax
\EndOfBibitem
\bibitem[Benvegnu and McConnell(1993)]{benvegnu93}
D.~J. Benvegnu and H.~M. McConnell, \emph{J. Phys. Chem.}, 1993, \textbf{97},
  6686--6691\relax
\mciteBstWouldAddEndPuncttrue
\mciteSetBstMidEndSepPunct{\mcitedefaultmidpunct}
{\mcitedefaultendpunct}{\mcitedefaultseppunct}\relax
\EndOfBibitem
\bibitem[Cseh and Benz(1999)]{cseh99}
R.~Cseh and R.~Benz, \emph{Biophys. J.}, 1999, \textbf{77}, 1477--1488\relax
\mciteBstWouldAddEndPuncttrue
\mciteSetBstMidEndSepPunct{\mcitedefaultmidpunct}
{\mcitedefaultendpunct}{\mcitedefaultseppunct}\relax
\EndOfBibitem
\bibitem[Starke-Peterkovic \emph{et~al.}(2006)Starke-Peterkovic, Turner, Vitha,
  Waller, Hibbs, and Clarke]{starke06}
T.~Starke-Peterkovic, N.~Turner, M.~F. Vitha, M.~P. Waller, D.~E. Hibbs and
  R.~J. Clarke, \emph{Biophys. J.}, 2006, \textbf{90}, 4060--4070\relax
\mciteBstWouldAddEndPuncttrue
\mciteSetBstMidEndSepPunct{\mcitedefaultmidpunct}
{\mcitedefaultendpunct}{\mcitedefaultseppunct}\relax
\EndOfBibitem
\bibitem[Wilke \emph{et~al.}(2006)Wilke, Dassie, Leiva, and Maggio]{wilke06}
N.~Wilke, S.~A. Dassie, E.~P.~M. Leiva and B.~Maggio, \emph{Langmuir}, 2006,
  \textbf{22}, 9664--9670\relax
\mciteBstWouldAddEndPuncttrue
\mciteSetBstMidEndSepPunct{\mcitedefaultmidpunct}
{\mcitedefaultendpunct}{\mcitedefaultseppunct}\relax
\EndOfBibitem
\bibitem[Pond \emph{et~al.}(2011)Pond, Errington, and Truskett]{pond11}
M.~J. Pond, J.~R. Errington and T.~M. Truskett, \emph{J. Chem. Phys.}, 2011,
  \textbf{135}, 124513--1--9\relax
\mciteBstWouldAddEndPuncttrue
\mciteSetBstMidEndSepPunct{\mcitedefaultmidpunct}
{\mcitedefaultendpunct}{\mcitedefaultseppunct}\relax
\EndOfBibitem
\end{mcitethebibliography}
\bibliographystyle{rsc} 

\end{document}